\documentstyle[epsfig,11pt]{article}

\def\beq{\begin{equation}}
\def\eeq{\end{equation}}

\def\tk{\tilde{k}}

\title{\bf Ridge from strings}
\author{M.A.Braun$^a$, C.Pajares$^b$, V.V.Vechernin$^a$\\
$^a$ Dep. of High Energy physics,
 Saint-Petersburg State University, Russia\\
$^b$ Dep. of Particles, University of Santiago de Compostela, Spain}

\begin{document}
\maketitle

{\bf Abstract}
In the colour string picture with fusion and percolation it is shown that
long range azimuthal-rapidity correlations (ridge) can arise from the
superposition of many events with  exchange of clusters of different number of strings
and not from a single event. Relation of the ridge with the flow harmonics coefficients
is derived. By direct Monte-Carlo simulations, in the technique
previously used to calculate these coefficients,
ridge correlations are calculated for AA, pA and pp
collisions. The azimuthal anisotropy follows from the assumed quenching of the emitted particles in the strong colour fields inside string clusters.
It is confirmed that in pp collisions the ridge structure  only appears
in rare events with abnormally high multiplicity.  Comarison
with the experimental data shows a good agreement. Also a good agreement
is found for pPb collisions. For AA collisions a reasonable agreement
is found for both near-side
and away-side angular correlations although  it worsens
at intermediate angles.

\section{Introduction}
One of the most impressing discoveries at LHC is observation of
strong long-range rapidity $y$ correlations collimated at small
relative azimuthal angles $\phi$ in particle production in
proton-proton ~\cite{CMS} and proton-nucleus collisions
 ~\cite{CMS1,ALICE1,ATLAS1},
the so called "ridge". Before the similar
effect was discovered in nucleus-nucleus collisins
~\cite{PHOBOS,STAR,ALICE}.

Several approaches tried to understand this effect \cite{ref8}-\cite{ref16}.
Long-range rapidity correlations have been successfully described
since long ago (see e.g. \cite{ref17}, \cite{ref18}). It is more difficult
to explain mixed rapidity-azimutal angle correlations.
In the Colour-Glass Condensate approach to explain it specific diagrams
were studied corresponding to subdominant contributions at large
number of colours which may generate such $y,\phi$ correlations
~\cite{ref8}-\cite{ref13}.
In this approach ridge is related to correlations at initial stages
of particle production and is a property of basic emission process,
which is then translated in the observable picture
by the hydrodynamic flow. In \cite{ref16} the color field is assumed to be
distrbuted in domains in the transverse plane, the direction  the field different in different domains. Partons interacting with each domain remember this direction, which is the reason of angular anisotropy.

In this paper we study correlations in the colour string approach.
Colour strings picture has been able to successfully explain many observable
phenomena in the soft dynamics domain. One expects it to be also applicable
to the ridge problem in so far as one is dealing with relatively modest
transverse momenta.
One of the main advantage of the colour string model is that it
allows to consder both AA and proton-proton collisions on the same
footing ~\cite{pajares} and so gives a unified picture for the ridge
in different processes.

Existence of long-range rapidity correlations has long been known
in this model and was, in fact, one of its most spectacular
predictions. However in the previous simple string models azimuthal
dependence was not generated, so that the arising correlations
were flat in $\phi$. Recently a modification of the colour string model
was proposed in which azimuthal dependence is included on the event-by-event
basis ~\cite{BP,ref21}. This gives some hope to find ridge in the
colour string picture. Here we demonstrate that this hope is correct:
we calculate the $y,\phi$ correlations in the framework of
~\cite{ref21} and show that they indeed have the ridge form.
Note that similar findings were earlier reported in ~\cite{bautista}
in a simplified analyitical approach.

Note that, unlike ~\cite{ref11,ref12}, in the string approach
correlations follow not from the
basic emission process in a single event but rather from the
distribution of events in actual observations. In principle emission
from a single string (and thus in an event) could also
lead to non-trivial $y,\phi$ correlation in the spectra due to recoil
effect in the transverse space after the first emission ~\cite{nestor}.
However we show that this effect is small and fast dying with rapidity
distance, so that it cannot lead to long-range rapidity correlations.
This point is discussed in  section 3 after we
briefly discuss our
modified string picture, which introduces non-trivial $\phi$-dependence
on the event-by-event basis in Section 2.
Section 4 is devoted to the general discussion of the long rapidity
correlations in the colour string picture and serves as an
introduction to this problem. Section 5 is devoted to the
derivation of the  expression for the correlation coefficientin this
picture. Section 6 presents our numerical results for AA, pp and pA
collisions.
Some conclusions are collected in Section 6.

\section{String picture}

The colour string model was proposed some time ago to describe
multiparticle production in the soft region. Its basic ideas can
be found in original papers and in a review ~\cite{capella,kaidalov,brapaj2}.
Its application to the flow problem was developed in our previous paper
~\cite{ref21}. Here we only reproduce the main
points necessary to understand the technique.
It is assumed that in a high-energy collision between the partons of the
participants colour strings are stretched, which may be visualized as
a sequence of $q\bar{q}$ pairs created from the vacuum or alternatively as
a strong chromoelectric field generated by the participant partons.
The strings are assumed to possess a certain finite dimension in
the transvesrse space related to confinement.
Each string then
breaks down in parts several times until its energy becomes of the order
of several GeV and it becomes an observed hadron. The number of strings
in the interaction area depends on the total available energy and
partonic structure of the colliding particles:
it grows with energy and atomic number. When the number of strings is
small they occupy a small part of the whole interaction area like drops
of liquid at considerable distance from one another. However when the number
of string grows they begin to overlap and fuse giving rise to strings with
more colour and covering more space in the interaction area. At a certain
critical string density strings begin to fuse forming clusters
 of the dimension comparable to that of the
interaction area (string percolation).
The basic assumptions which lie at the basis of the colour string picture
are supported by its very successful application to
multiparticle production in the soft region. It describes
well  the multiplicity and transverse momentum distributions
and many other details of the particle spectra.
The colour string picture has a certain similarity (see ~\cite{dddp})
with the saturation
(Colour Glass Condensate or Glasma) models, where the dynamics is
explained by the classical gluon field stretched
between the colliding hadrons. The effective number of
independent colour sources in string percolation can be put
in correspondence with the number of colour flux tubes
in the Glasma.
It is found that they indeed have the same energy and number of participants
dependence. As a consequence predictions of both approaches for most
of the observables are similar.

It is assumed that strings
decay into particles ($q\bar{q}$ pairs) by the well-known mechanism for
pair creation in a strong
electromagnetic field. In its simplest version,
the particle distribution at the moment of its production by the string is
\beq
P(p,\phi)=Ce^{-\frac{p_0^2}{T}}.
\label{prob}
\eeq
where $p_0$ is the particle initial transverse momentum,
$T$ is the string tension (up to an irrelevant numerical coeffcient) and
$C$ is the normalzation factor.
However, as proposed in ~\cite{ref21}, $p_0$ is different from the
observed particle momentum $p$ because
the particle has to pass through the  fused string area and emit gluons
on its way out. So in fact in Eq. (\ref{prob}) one has to consider $p_0$
as a function of $p$ and path length $l$ inside the nuclear overlap:
$p_0=f(p,l(\phi))$ where $\phi$ is the azimuthal angle. Note that Eq. (\ref{prob})
describes the spectra only at very soft $p_0$. To extend its validity to
higher momenta one may use the idea that the string tension fluctuates, which
transforms the Gaussian distribution into the thermal one ~\cite{bialas,deupaj}:
\beq
P(p,\phi)=Ce^{-\frac{p_0}{\sqrt{T/2}}}.
\label{probb}
\eeq
To describe the energy loss of the parton due to gluon emission
one may use the corresponding QED picture for a charged particle  moving
in the external electromagnetic field \cite{nikishov}. This leads to the
the quenching formula ~\cite{ref21}
\beq
p_0(p,l)=p\Big(1+\kappa p^{-1/3}T^{2/3}l\Big)^3,
\label{quench1}
\eeq
with the quenching coefficient $\kappa$ to be taken from the experimental
data. We adjusted $\kappa$  to
give the experimental value for the coefficient $v_2$
in mid-central Au+Au collisions at 200 GeV, integrated over the
transverse momenta.

Of course the possibility to use electrodynamic formulas for
the chromodynamic case may raise certain doubts. However in
~\cite{mikhailov} it was found that at least in the $N=4$ SUSY
Yang-Mills case the loss of energy of a coloured charge moving
in the external chromodynamic field was given by essentially the same
expression as in the QED.

\section{Correlations in emissions from a single string}
\subsection{Invariant production probability}
Consider a string stretched between two partons (e.g. quark and diquark)
with momenta $p_1$ and $p_2$, $p_1^2=p_2^2=0$.
We introduce two  orthogonal momenta $p$ and $q$ which form the plane
orthogonal to emission momenta:
\[ p=p_1+p_2,\ \ q=p_1-p_2,\ \ (pq)=0.\]
To define the emission probability for a parton of momentum $k$ we
introduce momentum $\tilde{k}$ orthogonal to plane $p,q$
\beq
\tk=k-p\frac{(kp)}{p^2}-q\frac{(kq)}{q^2}.
\label{ortho}
\eeq
We trivially find
\[(p\tk)=(q\tk)=0\]
and
\beq
\tk^2=k^2-\frac{(kp)^2}{p^2}-\frac{(kq)^2}{q^2}.
\label{square}
\eeq
Let $p_{1\perp}=p_{2\perp}=0$. Then $p_{1-}=p_{2+}=0$, so that
\[ p_+=p_{1+},\ \ p_-=p_{2-},\ \ q_+=p_+,\ \ q_-=-p_-\]
and we find
\[
\tk_+=k_+-p_+\frac{k_+p_-+k_-p_+}{2p_+p_-}+
p_+\frac{-k_+p_-+k_-p_+}{2p_+p_-}=0,\]
\[
\tk_-=k_+-p_-\frac{k_+p_-+k_-p_+}{2p_+p_-}-
p_-\frac{-k_+p_-+k_-p_+}{2p_+p_-}=0,\]
\[ \tk_\perp=k_\perp,\]
as expected.
So we can take for the invariant density
(suppressing the normaization factor)
\beq
\rho(k)=\frac{dW}{dyd^2k_\perp}=e^{a\tk^2},
\eeq
which in the  case $p_{1\perp}=p_{2\perp}=0$ transforms into the
standard expression
\beq
\rho({\bf k})=e^{-a{\bf k}^2}.
\eeq

\subsection{Double emission}
As mentioned, the string picture is oriented towards the soft dynamics.
Simultaneous emission from the string of two jets with large rapidity
difference is assumed to be strongly damped. Its probablity
for rapidity difference
$\Delta y=y_1-y_2>0$  is assumed
to be proportional to
$\exp (-aM^2)\delta^2(k_1+k_2)$ where $M^2=k_{1\perp}k_{2\perp}e^{\Delta y}
-2(k_1k_2)_\perp$ is the total mass
So it gives strong
back-to-back azimuthal correlations, as for hard emissions, but they
fall as $\exp(-\Delta y)$.

Long-range rapidity correlations in
emission from the string  may rather follow from their sequential decay
into two partons.

Let us assume that after the first emission of a particle with momentum
$k_1$ this particle forms a new string with the target from which a second
emission follows producing a second particle with momentum $k_2$.
The total probability is obviously given by the product
\beq
\rho_1(k_1)\rho_2(k_1,k_2).
\eeq
The initial probability is
\beq
\rho_1(k_1)=e^{-a{\bf k}_1^2}
\eeq
and the second one is
\beq
\rho_2{k_2}=e^{a\tk_2^2},
\eeq
where $\tk_2$ is orthogonal to plane $k_1,p_2$ and given by (\ref{ortho})
with $p=k_1+p_2$ and $q=k_1-p_2$.
We have
\[(k_2p)=k_{2+}k_{1-}+k_{2-}k_{1+}+(k_2k_1)_\perp+k_{2+}p_{2-},\]
\[(k_2q)=k_{2+}k_{1-}+k_{2-}k_{1+}+(k_2k_1)_\perp-k_{2+}p_{2-},\]
\[ p^2=2k_{1+}p_{2-}=-q^2.\]
So
\[\tk_2^2=k_2^2-\frac{(k_2p)^2-(k_2q)^2}{p^2}=
k_2^2-\frac{(k_2,p+q)(k_2,p-q)}{p^2}.\]
Calculating this we have
\[\tk_2^2=k_2^2-2\frac{k_{2+}}{k_{1+}}
(k_{2+}k_{1-}+k_{2-}k_{1+}+(k_2k_1)_\perp)=
k_{2\perp}^2-2\frac{k_{2+}}{k_{1+}}(k_2k_1)_\perp)+
\frac{k_{2+}^2}{k_{1+}^2}k_{1\perp}^2=
\Big(k_{2}-\frac{k_{2+}}{k_{1+}}k_1\Big)_\perp^2.\]
So we find
\beq
\rho_2(k_2)=e^{a\Big(k_{2}-\frac{k_{2+}}{k_{1+}}k_1\Big)_\perp^2},
\eeq
or in Euclidean momenta and rapidities
\beq
\rho_2(y_1,y_2,{\bf k}_1,{\bf k}_2)=
e^{-a\Big({\bf k}_{2}-e^{-\Delta y}{\bf k}_1)^2}.
\eeq

The total probability for the double production expressed in Euclidean
transverse momenta $k_1$ and $k_2$ becomes
\beq
\frac{dW}{dy_1dy_2d^2k_1d^2k_2}=
e^{-a(q_1^2+q_2^2+q_1^2e^{-2 \Delta y}-2q_1q_2\cos\phi e^{-\Delta y}}.
\eeq
Recall that we assume $\Delta y=y_1-y_2>0$

The first two terms independent of the relative rapidity correspond
to independent emission. The rest depend on the rapidity difference
and exponentially fall with it. Correlations are contained in the last
term. Unfortunately they again fall like $\exp(-\Delta y)$ with the
rapidity distance between emitted particles.

So our conclusion is that the correlations in emissions from a
single string are all of the short-range character in rapidity.

\section{Long-range correlations}
\subsection{Correlations for given string configuration}
Consider  an event in which $n$ strings are formed, which may be different.
The difference may come both from the color of strings (simple or fused)
and from their different location in the overlap area. The inclusive cross-
section from the $i$-th string is $I_i(y,\phi)$. We normalize it as follows
\beq
\int dyd\phi I_i(y,\phi)=<N_i>,
\label{norm}
\eeq
where $<N_i>$ is the average number of particles emitted from string $i$.
The total inclusive cross-section is
\beq I(y,\phi)=\sum_{i=1}^nI_i(y,\phi).
\label{incltot}
\eeq
Integration over $y,\phi$ gives the total mean number of particles,
$<N>=\sum_i<N_i>$, emitted by the given string configuration.
Note than the contribution to the sum in (\ref{incltot}) only comes
from the strings which cover rapidity $y$. Namely, if the upper and lower
ends of
the string $i$  have rapidities $y_i^u$ and $y_i^l$ respectively
then the contributing strings must have $y_i^l<y<y_i^u$.

Now consider the double inclusive cross-section at azimuthal angles
$\phi_1$ and $\phi_2$ and  rapidities $y_1$ and $y_2$.
We can divide all strings into three groups. Strings in the first
group cover only rapidity $y_1$. String in the second group cover only
rapidity $y_2$. Finally strings in the third group cover both
rapidities $y_1$ and $y_2$. The total inclusive cross-section at
$y_1$ will be given by (\ref{incltot}) with the sum including strings
of the first and third group:
\beq I(y_1,\phi_1)=\sum_{i}\Big(I^{(1)}_i(y_1,\phi_1)+
I^{(3)}_i(y_1,\phi_1)\Big).
\label{incl1}
\eeq
At $y_2$ the inclusive cross-section will come from strings of the
second and third group:
\beq I(y_2,\phi_2)=\sum_{i}\Big(I^{(2)}_i(y_2,\phi_2)+
I^{(3)}_i(y_2,\phi_2)\Big),
\label{incl2}
\eeq
Two particles at rapidities and angles $y_1,\phi_1$ and $y_2,\phi_2$
may come either from different strings or from the same string.
In the first case the double inclusive cross-section will be given by
the expression
\[
I_1(y_1,\phi_1,y_2,\phi_2)=\sum_{i,k}\Big(I^{(1)}_i(y_1,\phi_1)
I^{(2)}_k(y_2,\phi_2)+I^{(1)}_i(y_1,\phi_1)I^{(3)}_k(y_2,\phi_2)\]\beq+
I^{(2)}_i(y_2,\phi_2)I^{(3)}_k(y_1,\phi_1)\Big)
+\sum_{i\neq k}I^{(3)}_i(y_1,\phi_1)I^{(3)}_k(y_2,\phi_2).
\label{i1}
\eeq
In the last term the two particles come from
different strings of the third group.
In the second case both particles come from the same string of the
third group. If we neglect correlations inside the string following
the results of the preceding section then this contribution will be
\beq
I_2(y_1,\phi_1,y_2,\phi_2)=
\sum_iI^{(3)}_i(y_1,\phi_1)I^{(3)}_i(y_2,\phi_2).
\label{i2}
\eeq
Here we have to make an important comment. Eq. (\ref{i2}) is true if
the number of emitted particles is equal or greater than two. If it is
equal to 1 then this contribution does not exist.

Neglecting this rare possibility
and summing these two parts we get the total double inclusive
cross-section as
\[
I(y_1,\phi_1,y_2,\phi_2)=\sum_{i,k}\Big(I^{(1)}_i(y_1,\phi_1)+
I^{(3)}_k(y_2,\phi_2)\Big)\Big(I^{(12)}_i(y_2,\phi_2)+
I^{(3)}_k(y_1,\phi_1)\Big)\]\beq
=I(y_1,\phi_1)I(y_2,\phi_2).
\label{itot}
\eeq
This means that there are no correlations for events
with the same string configuration.

Note that from Eq. (\ref{itot}) it follows that
\beq
J\equiv \int dy_1d\phi_1dy_2d\phi_2 I(y_1,\phi_1,y_2,\phi_2)=<N>^2.
\label{jnorm}
\eeq
In fact $J=<N(N-1>=<N^2>-<N>$
But for the Poisson distribution of the particles emitted from strings
$<N^2>=<N>^2+<N>$ in accordance with (\ref{jnorm}).

\subsection{Long-range correlations from fluctuating string configuration}
As follows from the previous considerations, in the colour string
picture correlations can arise from the superposition of many events with
different number and type of strings. In fact appearence of long-range
correlations in this picture was observed long ago and was considered as
one of the main consequences of the string picture.

The simplest type of correlations, long discussed in literature, are
the forward-backward correlations relating the probability to observe
particles in the backward rapidity window with a given number of particles
in the forward rapidity window. Appearence of such correlations is evident
from the following reasoning. Let all the strings be equal for simplicity,
If the number of strings may be
different in different events, then in an event with, say, $n$
strings the number of particles observed in the forward rapidity window is
$n$ times greater than from a single string. But in this event also the
number of particles observed in the backward rapidity window will be
$n$ times greater than from a single string, so that an obvious correlation
follows. This argument was later generalized to fusion and percolation of
strings.

However passing to the azimuthal angle dependence one concludes that if
emission from strings is isotropic, independent of their type,
the correlations due to their distribution in different events will also be
isotropic. Also in the central rapidity region the inclusive
cross-sections are practically independent of rapidity. This generates
a plateau in the $\delta y-\delta\phi$ distribution
rather than a ridge, with only a narrow peak  at small $\delta\phi$ and
$\delta y$ due to short range correlations.

This conclusion remains true if one averages the inclusive cross-sections
over all events with the resulting loss of azimuthal angle dependence.
So the ridge can only be obtianed on the event-by-event basis.

\section{Ridge and the flow coefficients}
As discussed above the ridge in our picture  arises due to
fluctuations in
both string distributions and impact parameter.
So it is important to study
the formation of averages relevant to the ridge.

For a particular string configuration with a fixed azimuthal angle $\phi_0$ of the
impact parameter
the inclusive cross-section is found as a function of angle $\phi$ as
\[
I^c(y,\phi)=A^c(y)+2\sum_{n=1}\Big(B_n^c(y)\cos n(\phi-\phi_0)+
C^c_n(y)\sin n(\phi-\phi_0)\Big)\]\beq
=
A^c\Big(1+2\sum_{n=1}\Big(b_n^c(y)\cos n\phi+
c_n^c(y)\sin n\phi\Big).
\label{eq1}
\eeq
The flow coefficients for a given event are given by
\beq
v_n^c(y)=\Big((a_n^c(y))^2+(b_n^c(y))^2\Big)^{1/2}.
\label{vn}
\eeq
The experimentally observed flow coefficitnts are obtained after
averaging over different
string distributions, which we denote as $<...>$
\beq
v_n(y)=<v_n(y)>=<\Big((a_n(y))^2+(b_n(y))^2\Big)^{1/2}>.
\label{vnexp}
\eeq

Passing to correlations, for a given string configuration we have the double
inclusive cross-section given by
Eq. (\ref{itot}):
\[
I^{e}(y_1,\phi_1,y_2,\phi_2)\]\[=\Big(A^c(y_1)+2\sum_{n=1}
\Big(B_n^c(y_1)\cos n(\phi_1-\phi_0)+
C_n^c\sin n(\phi_1-\phi_0)\Big)\]\beq
\times
\Big(A^c(y_2)+2\sum_{m=1}
\Big(B_m^c(y_2)\cos m(\phi_2-\phi_0)+
C_m^c(y_2)\sin m(\phi_2-\phi_0)\Big).
\label{eq2}
\eeq
We have to average this expression over string distributions and directions of the impact
parameter $\phi_0$. The latter reduces to integration over $\phi_0$
with weight $1/2\pi$.
Doing first this integration and then averaging over string distribution we obtain the
"experimental" double inclusive cross-section
\beq
I(y_1,y_2,\phi_{12})=<A(y_1)A(y_2)>+2\sum_n <W_n(y_1,y_2)>
\cos n\phi_{12},
\label{iexp}
\eeq
where $\phi_{12}=\phi_1-\phi_2$ and
\beq
W_n(y_1,y_2)=B_n(y_1)B_n(y_2)+C_n(y_1)C_n(y_2).
\eeq
Similar averaging of the single inclusive cross-section obviously
eliminates all oscillating terms,
so that
\beq
I(y_1)=<A(y_1)>.
\label{i1exp}
\eeq

Thus we find the correlation function as
\[
C(y_1,y_2,\phi_{12})=
\frac{<(A(y_1)A(y_2)>+2\sum_n<W_n(y_1,y_2)>\cos n\phi_{12}}
{<A^c>^2}-1\]\beq=
\frac{DA(y_1,y_2)}{<A(y_1><A(y_2)>}+2\sum_{n=1}w_n(y_1,y_2)\cos n\phi_{12},
\label{cor}
\eeq
where $DA(y_1,y_2)=<A(y_1)A(y_2)>-<A(y_1><A(y_2)>$ is the covariance
of the integrated single inclusive cross-sections,
that is  multiplicities, and
\beq
w_n(y_1,y_2)=\frac{<(W_n(y_1,y_2)>}{<A>^2}.
\eeq

As we  observe the correlation contains two terms. The first is due to
fluctuations in the total
multiplicity and is independent of the angle. The ridge comes from the
second term,
which depends on averages of the product of initial coefficients $B_n$
and $C_n$ at different rapidities.

As mentioned, in the central region with suffciently long strings
the distributions are practically independent of rapidity. This creates
a plateau in rapididity with the angular dependence of correlations given
by
\beq
C(\phi_{12})=
\frac{DA}{<A>^2}+2\sum_{n=1}w_n\cos n\phi_{12}
\label{cor1}
\eeq
and
\beq
DA=<A^2>-<A>^2,\ \ w_n=\frac{<(B_n)^2+(C_n)^2>}{<A>^2}.
\eeq

If one neglects
fluctuations in the multiplicity at a given centrality and assumes
$<A^2>=<A>^2$ then one finds
\beq
w_n=<(a_n)^2+(b_n)^2>=<(v_n)^2>,
\eeq
that is the average of the individual flow coefficients squared. This should
be compared with the normally defined flow coefficients Eq. (\ref{vnexp}),
which are obtained by averaging of individual flow coefficients themselves.
If fluctuations both in $A$  and in $v_n$ are neglegible  one can take
$<(v_n)^2>=<v_n>^2=v_n^2$ and thus
find the ridge directly from the flow coefficients.

\section{Numerical calculations}
\subsection{AA collisions}
The general scheme of calculations repeats the one presented in
our previous paper dedicated to flow coeficients ~\cite{ref21}.
So here we only briefly describe the main points. A Monte-Carlo code
was developed which first distributes the colliding nucleons in the
transverse area according to the nuclear profile functions
at a given impact parameter $b$. The number of interacting nucleons
(participants) is determined according to the Glauber picture.
Then strings are attached to participants. The number of strings per
participant for a given centrality is determined from the conclusions
of ~\cite{tarn} for energies 62.4 and 200 GeV and of ~\cite{DDD} for
the LHC energy of 2.76 TeV. Strings are assumed to fuse if they are
located in a common area of radius $r_s=0.32 fm$. The colour
and tension of the  string fused from $n$ original strings are
taken to be $\sqrt{n}$ greater than for the original string.
Particles are emitted from fused strings according to the thermal
distribution in transverse momentum.
The anisotropy of particle distribution is assumed to come from the
passage of particles through the gluon field inside the strings
and the corresponding quenching of their transverse momentum.
As mentioned, the concrete form of this quenching is borrowed from a
similar process in the quantum electrodynamics (see ~\cite{nikishov}).
As a result of this Monte-Carlo code one obtains single and double
inclusive cross-section in the forms (\ref{eq1}) and (\ref{eq2}).
Averaging gives the coefficients $v_n$ and $w_n$ and the correlation
coefficent $C(\phi)$.

Particle emission in the central region turns out to be
practically independent of rapidity. So the obtained correlation
coefficient corresponds to the ridge form while the rapidity
distance does not become comparable with the overall rapidity. At such
large rapidity distance one has to seriously take into account energy
conservation and the dependence on the two rapidities $y_1$ and $y_2$
in Eq. (\ref{cor}).

\begin{figure}
\hspace*{2.5 cm}
\epsfig{file=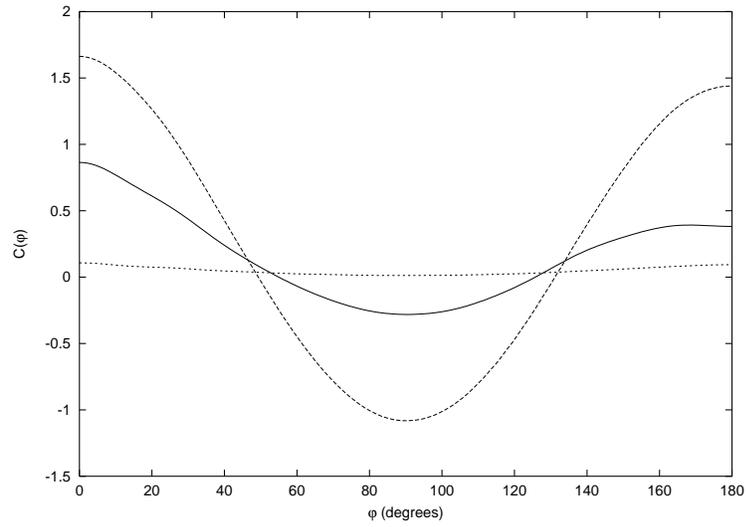,width=10 cm}
\caption{Correlation coefficients $C(\phi)$ for Au-Au
collisions at 62.4 GeV for central(middle curve at small angles)
mid-central (upper curve) and peripheral (lower curve) events}
\label{fig1}
\end{figure}

\begin{figure}
\hspace*{2.5 cm}
\epsfig{file=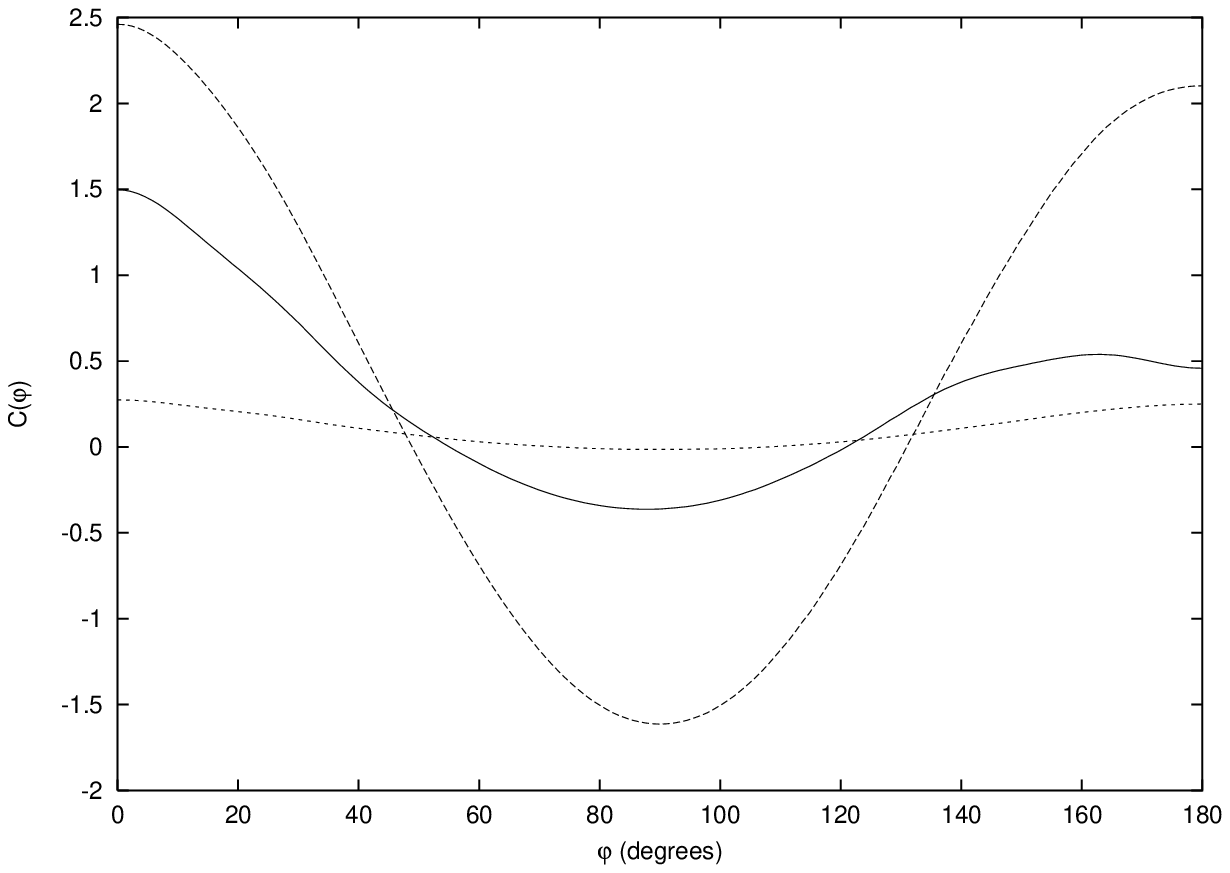,width=10 cm}
\caption{Correlation coefficients $C(\phi)$ for Au-Au
collisions at 200 GeV for central(middle curve at small angles)
mid-central (upper curve) and peripheral (lower curve) events}
\label{fig2}
\end{figure}

\begin{figure}
\hspace*{2.5 cm}
\epsfig{file=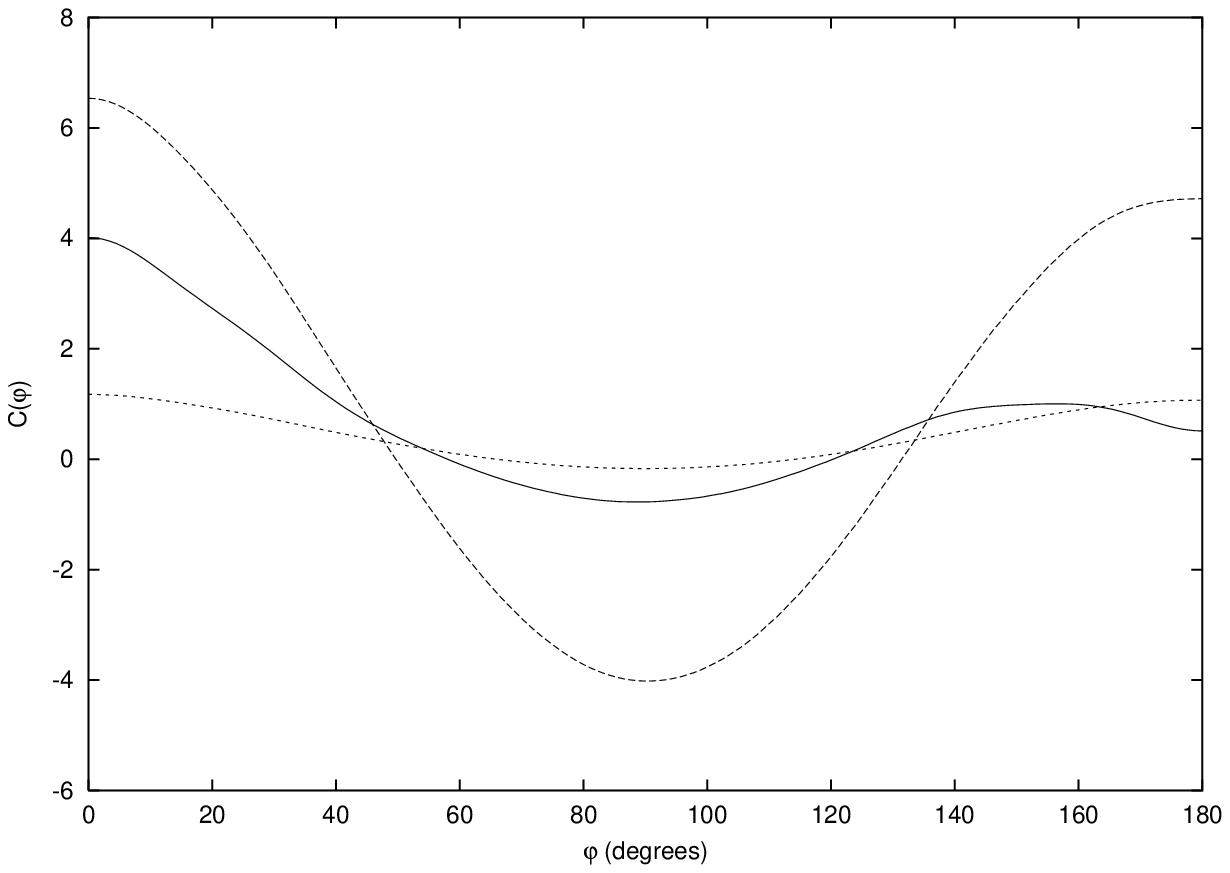,width=10 cm}
\caption{Correlation coefficients $C(\phi)$ for Pb-Pb
collisions at 2.76 TeV for central(middle curve at small angles)
mid-central (upper curve) and peripheral (lower curve) events}
\label{fig3}
\end{figure}

Using the constructed Monte-Carlo code we calculated
$w_n$ for $n=1,...16$ and then found coefficient $C(\phi)$
for energies 62.4, 200 GeV and 2.76 TeV and different centralities.
The experimental data at a given cenrality
mostly assume multipying $C(\phi)$ by the multiplicity minus unity:
\beq
 C(\phi)=\to <A> C(\phi).
\eeq
The resulting correlation coefficients for the
above mentioned three energies are presented in Figs. \ref{fig1}
-\ref{fig3}.
In thes figures,
as in the majority of the following ones, the constant term
$DA/<A>^2$ is dropped. For AA collisions it is small (of the order of 0.1)

\begin{figure}
\hspace*{2.5 cm}
\epsfig{file=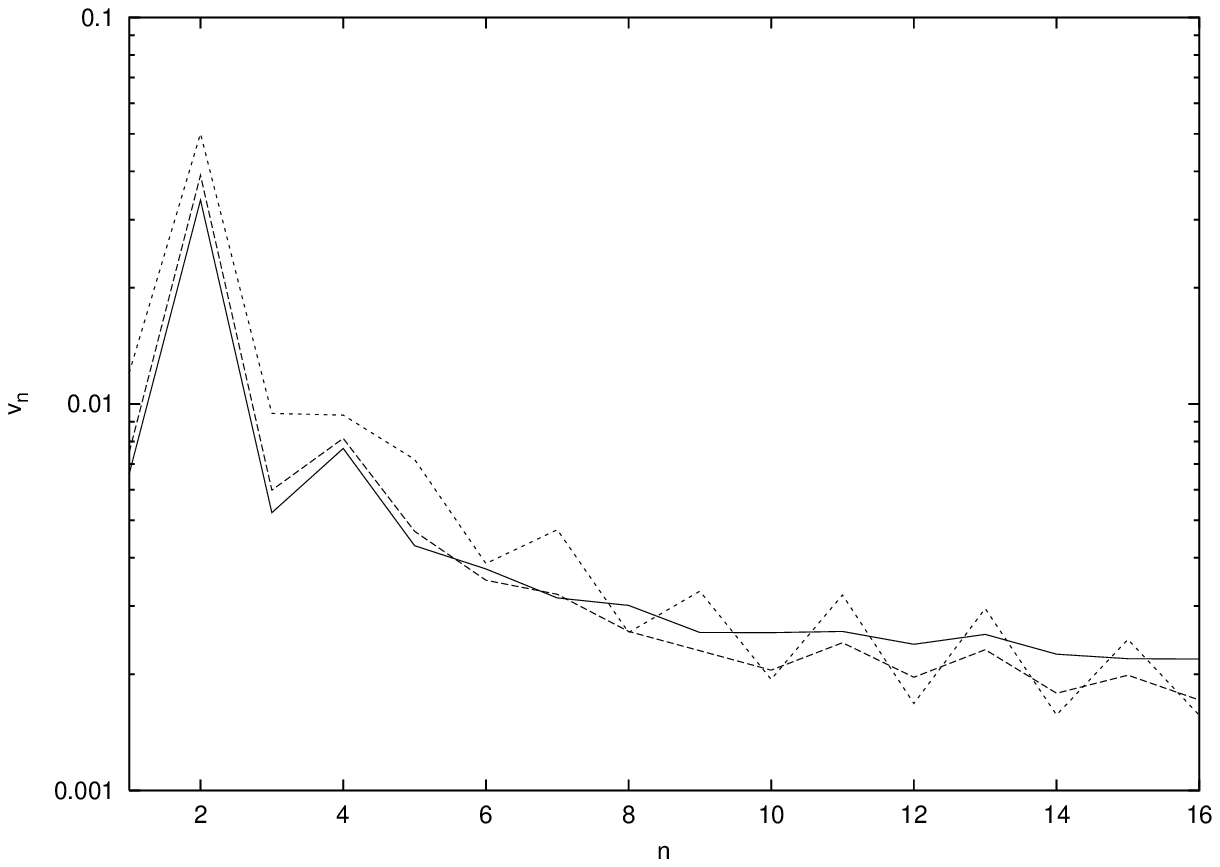,width=10 cm}
\caption{Flow coefficients $v_n$ for minimum bias events in Au-Au
collisions at 62.4 and 200 GeV
(lower and middle curves at $n=2$ respectively)
and in Pb-Pb collisions at 2.76 Tev (upper curve at $n=2$)}
\label{fig4}
\end{figure}

\begin{figure}
\hspace*{2.5 cm}
\epsfig{file=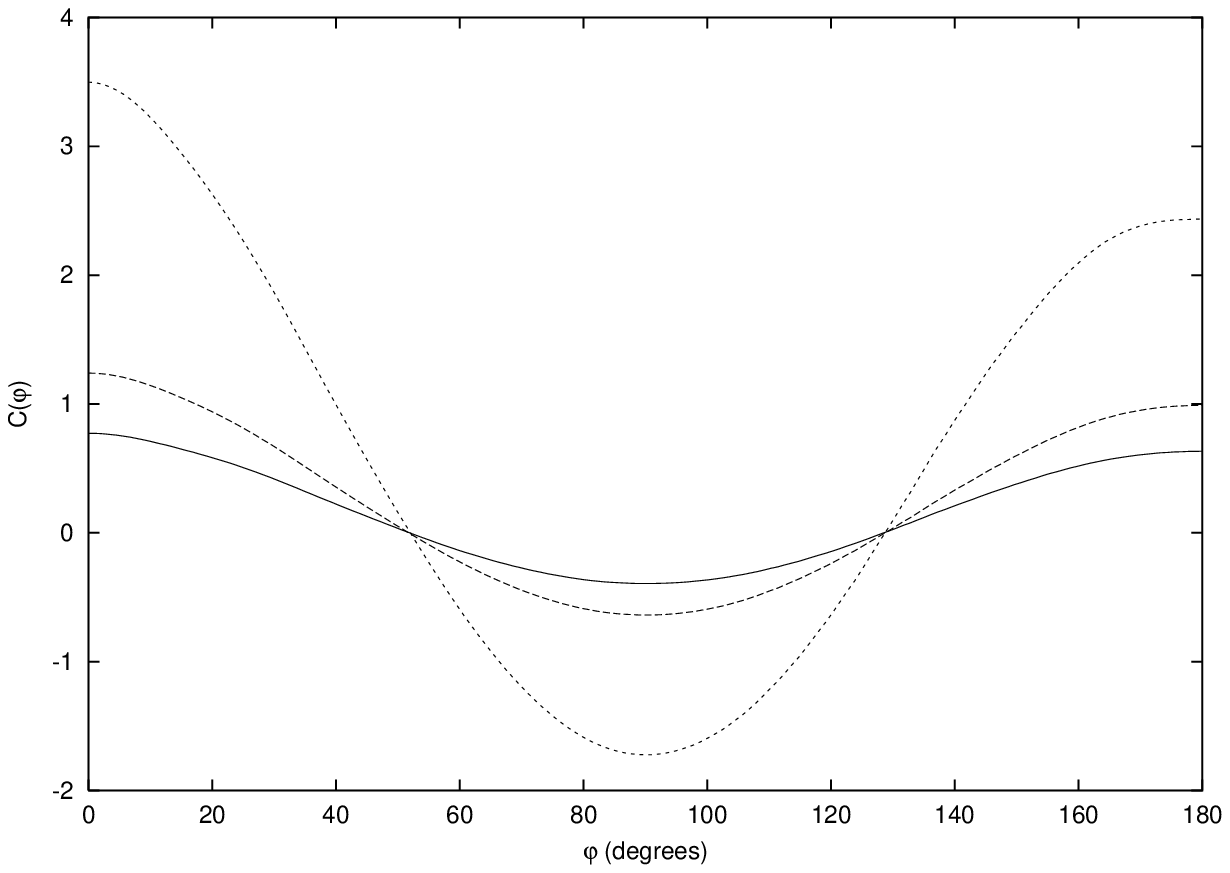,width=10 cm}
\caption{Correlation coeficients $C(\phi)$ for minimum bias
events in
Au-Au collisions at 62.4 GeV(upper curve at 90$^o$, 200 GeV
(middle curve at  90$^o$) and Pb-Pb
collisions at 2.76 TeV (lower curve at  90$^o$)}
\label{fig5}
\end{figure}

To illustrate the energy dependence we present the flow coefficients $v_n$
and correlation coefficients $C(\phi)$ for all the three energies for
minimum bias events in Figs. \ref{fig4} and \ref{fig5}.

\begin{figure}
\hspace*{2.5 cm}
\epsfig{file=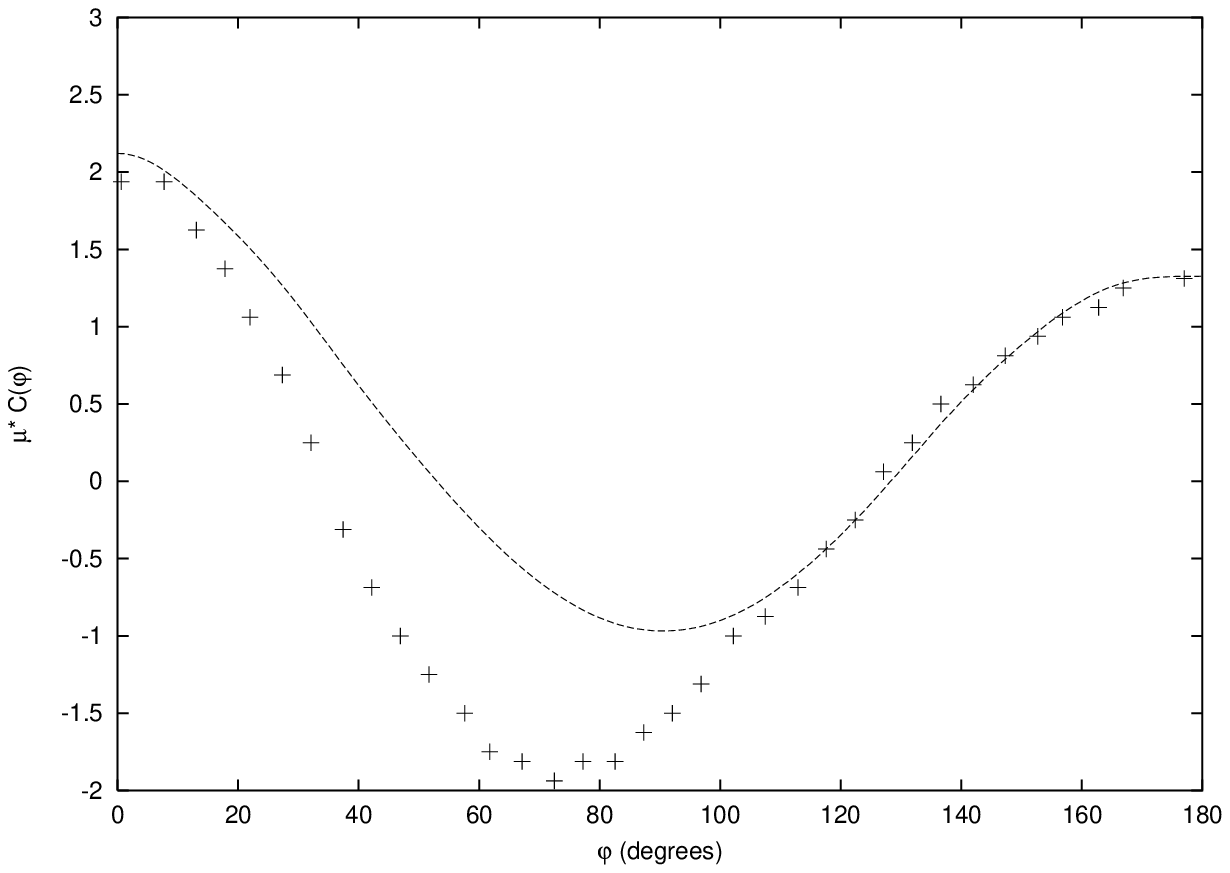,width=10 cm}
\caption{Correlation coefficient $C(\phi)$
for Au-Au at 200 GeV for 10\% of the most central events
against the experimental data from ~\cite{STAR}}
\label{fig6}
\end{figure}

We compared our results with the experimental data for
Au-Au colisions at 200 GeV presented in
~\cite{STAR}
in Fig. \ref{fig6} where we show the calculated
correlation coefficients
$C(\phi)$
for 10\% of the
most central events against the experimental data from
STAR.

\begin{figure}
\hspace*{2.5 cm}
\epsfig{file=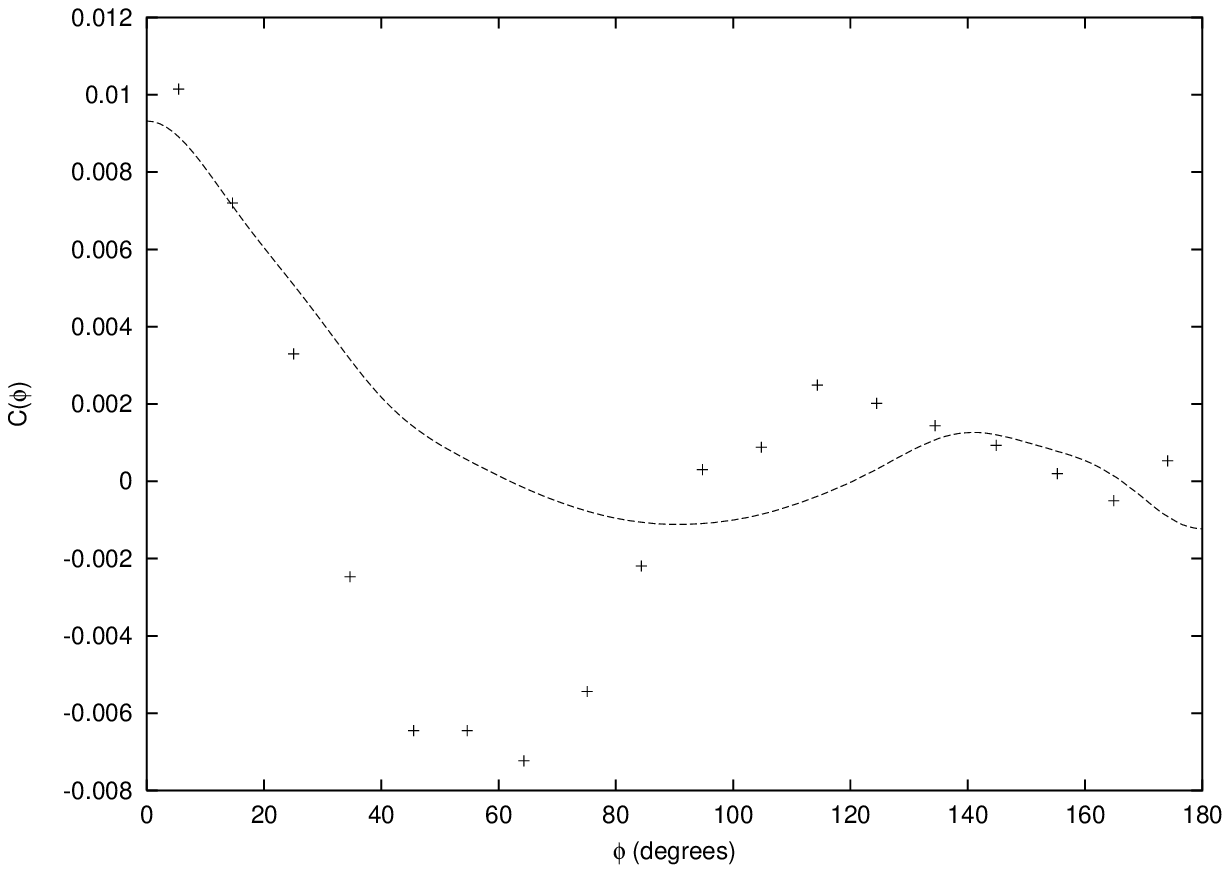,width=10 cm}
\caption{Correlation coefficient $C(\phi)$
for Pb-Pb at 2.76 TeV for the most central events
against the experimental data from ~\cite{ALICE1}}
\label{fig6a}
\end{figure}

\begin{figure}
\hspace*{2.5 cm}
\epsfig{file=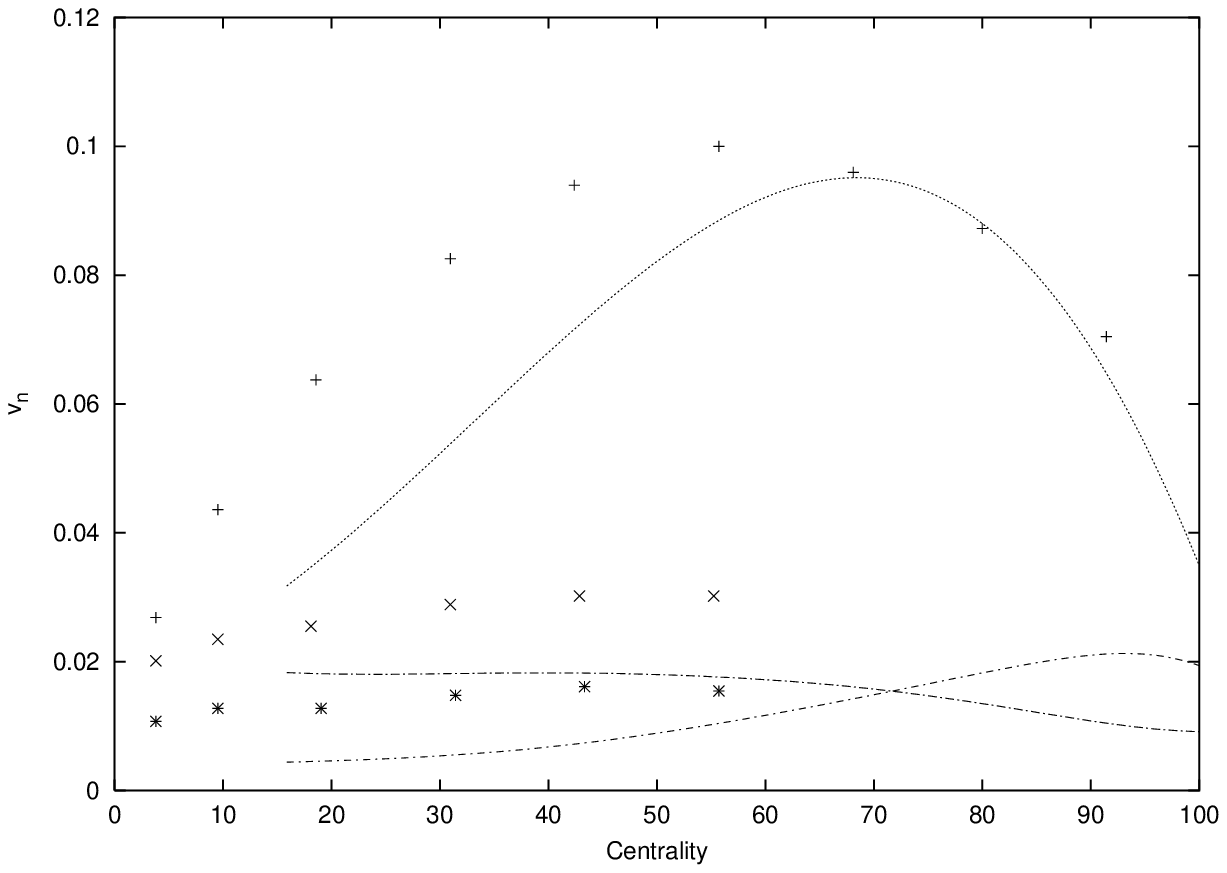,width=10 cm}
\caption{Flow coefficients $v_2$, $v_3$
and $v_4$ (from top to bottom on the left) at different centraliries
for Pb-Pb at 2.76 TeV
against the experimental data from ~\cite{ALICE}}
\label{fig6b}
\end{figure}

At the LHC energy 2.76 TeV we compared our results with the data
from ALICE ~\cite{ALICE1} at central collisions with detected pairs
of particles with rather large transverse momenta $p_T$ between 1 and 3 GeV/c.
To adjust to the average $p_T$ we raised the minimal value of $p_T$ for
our total cross-sections to 0.5 GeV/c. The resulting correlation
coefficients for most central collisions are compared wit ALICE data
in Fig. \ref{fig6a}
The corresponding flow coefficients at different centralities
are shown in Fig. \ref{fig6b} again
compared to ALICE data.

As we observe, in both cases the form of our results
is somewhat distorted
as compared to the experimental data.
We  obtain a  reasonable agreement with the
data at angles $\leq 40^o$ and very good agreement at angles $\geq 110^o$.
However the
the experimental data reach their minimal value around 70$^o$,
whereas the calculated values are minimal at ~90$^o$ and  from 40$^o$ to
100$^o$ our results lie substantially above the data.
 The reason of this
disagreement probably has to do with simplifications made in our Monte-Carlo
simulations. In fact the tension of a cluster of $n$ overlapping strings
is roughly $\sqrt{nS_n/S_1}$ where $S_n$ and $S_1$ are areas of the cluster
and simple string respectively. With that, forms of clusters with a given number of strings may be quite various. To make calculation feasible, in our Monte-Carlo
code, as in ref. \cite{ref21}, it was assumed that all clusters have the same size and form,
with the tension of each cluster just $\sqrt{n}$. It was previously shown
that such simplfication does not influence the bulk properties of the string
picture, such as multiplicities and transverse momentum distribution of
emitted particles. However it neglects fluctuations in the size and form of
clusters and thus may distort the azimuthal asymmetry.
This is especially significant for AA collisions where the total
number of strings and average cluster dimension are much larger than in
pp and pA collisions.  This fact was already noted in \cite{ref21} where we
found smaller values for higher harmonics $v_n$, $n\geq 3$ as compared to
the data for Au-Au collisions at RHIC. In pp and pA collisions
the average dimension of clusters is much smaller, which may lead to better
agreement with the data.

\subsection{Ridge in pA collisions}
\begin{figure}
\hspace*{2.5 cm}
\epsfig{file=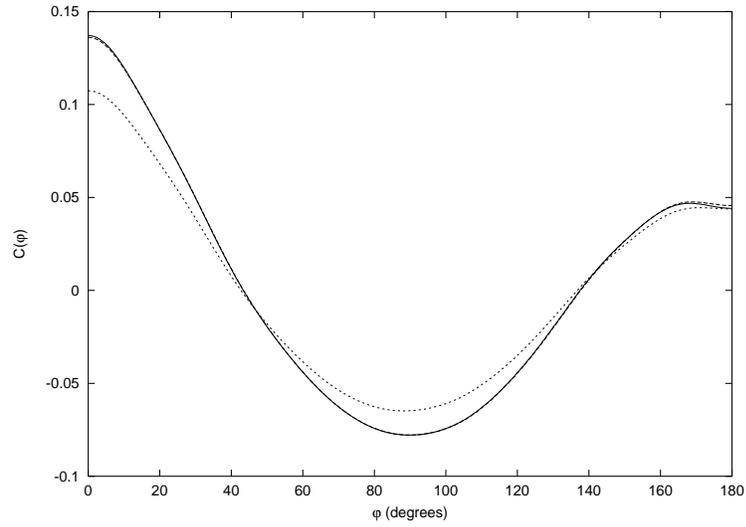,width=10 cm}
\caption{Correlation coeficients $C(\phi)$ for p-Pb
collisions at 5.02 TeV for central and mid-central
collisions (upper curve at small angles) and peripheral
collisions (lower curve)}
\label{fig11}
\end{figure}

In the string picture proton-nucleus collisions are described
in the similar manner as for nucleus-nucleus and proton-proton
collisions. The difference from the former case is that the
the strings are stretched between the projectile proton and all
nucleons of the target at a given impact parameter $b$.
We choose the maximal number of strings attached to the
nucleons of the target to be 18 at energies in the region 5-7 TeV,
in accordance with our results for
the multiplicity in proton-protons collisions. The number of strings
attached to the
projectile proton will correspondingly be $A^{1/3}$ times larger.
One might expect stronger dependence on the rapidity distance
due to asymmetry between the projectile and target. However
our calculations show that at least for $y_1-y_2\leq 4$ the
results remain independent of rapidity in the central region.
The resulting $w_n$ and $C(\phi)$ are similar for energies
62.4, 200 GeV and 5.02 TeV. So we limit ourselves by presenting
only our results for $C(\phi)$ at 5.02 TeV in Fig. \ref{fig11}.
As we observe $C(\phi)$ practically do not change with centrality
up to sufficiently peripheral events.

\begin{figure}
\hspace*{2.5 cm}
\epsfig{file=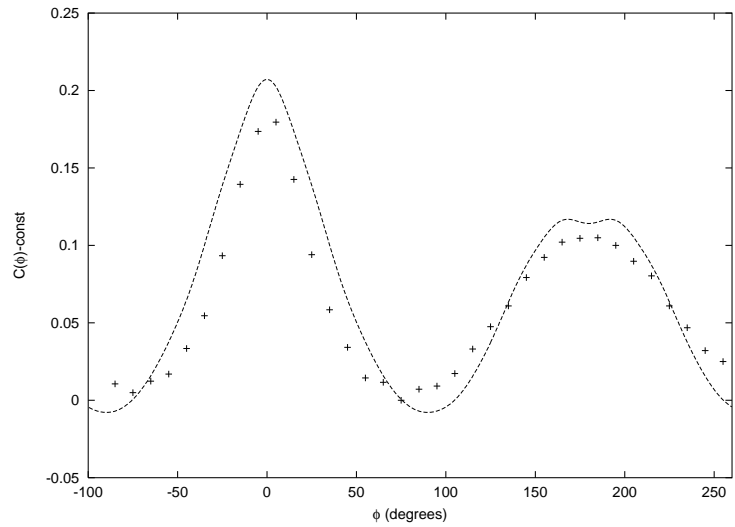,width=10 cm}
\caption{Correlation coeficient $C(\phi)$ for p-Pb
collisions at 5.02 TeV for central collisions compared
to the data in ~\cite{ALICE1} (with the ZYAM procedure).}
\label{fig11a}
\end{figure}

\begin{figure}
\hspace*{2.5 cm}
\epsfig{file=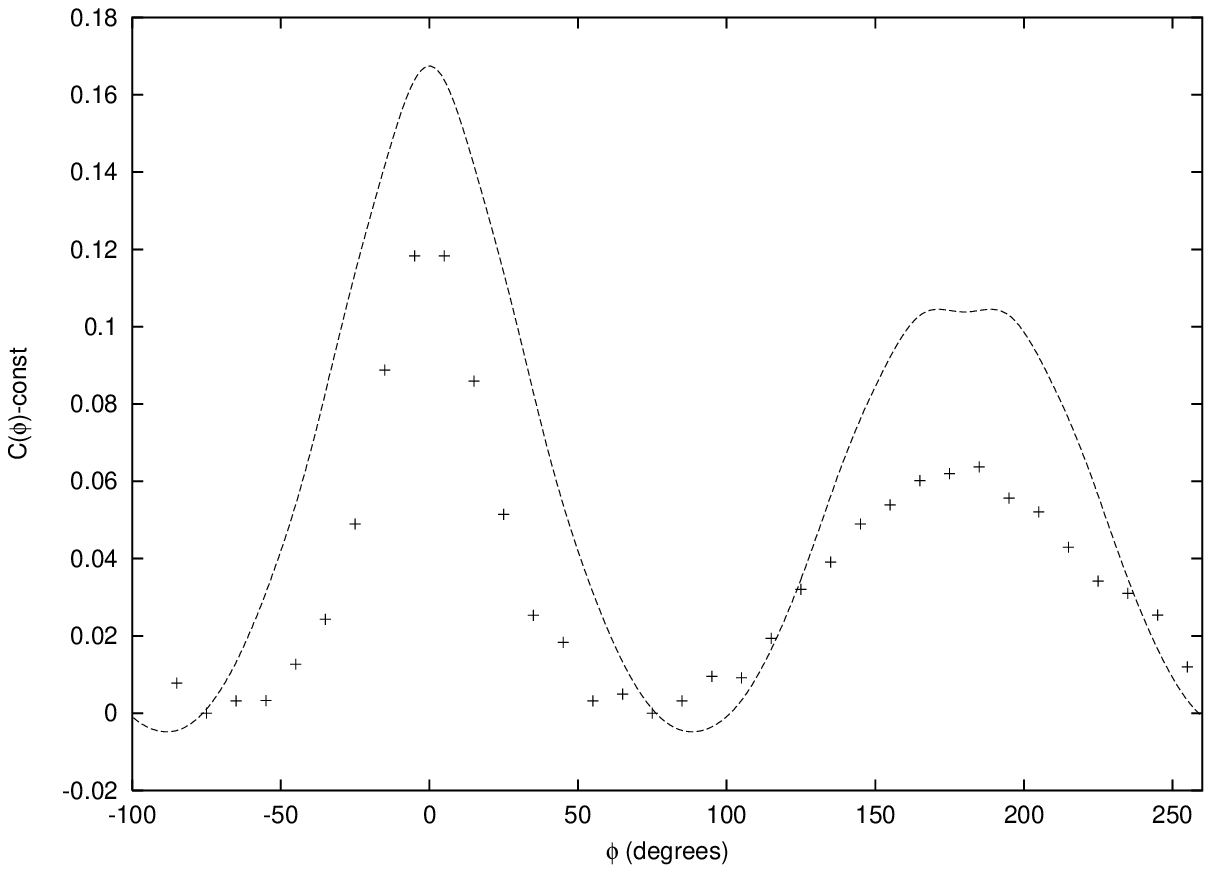,width=10 cm}
\caption{Correlation coeficient $C(\phi)$ for p-Pb
collisions at 5.02 TeV for peripheral collisions compared
to the data in ~\cite{ALICE1} (with the ZYAM procedure).}
\label{fig11b}
\end{figure}

We compared our results with the data from ~\cite{ALICE1} at
central and peripheral collisions in Figs. \ref{fig11a} and \ref{fig11b}.
with the use of the ZYAM (zero-yield-at-minimum) procedure.
The agreement at central collisions is quite good.
At peripheral colisions our results somewhat overshoot the data.
However they
agree with the tendency to have smaller $C(\phi)$ at less centrality.
Also our theoretical definition of centrality (the ratio of the
impact paramter to its maximal value) is different from the one used
in the experiment, so that our peripheral collisions are strongly
contaminated by the more central collisins  from the experimental
point of view, which may explain  comparatively large values of
$C(\phi)$.
\subsection{Ridge in pp collisions}
\begin{figure}
\hspace*{2.5 cm}
\epsfig{file=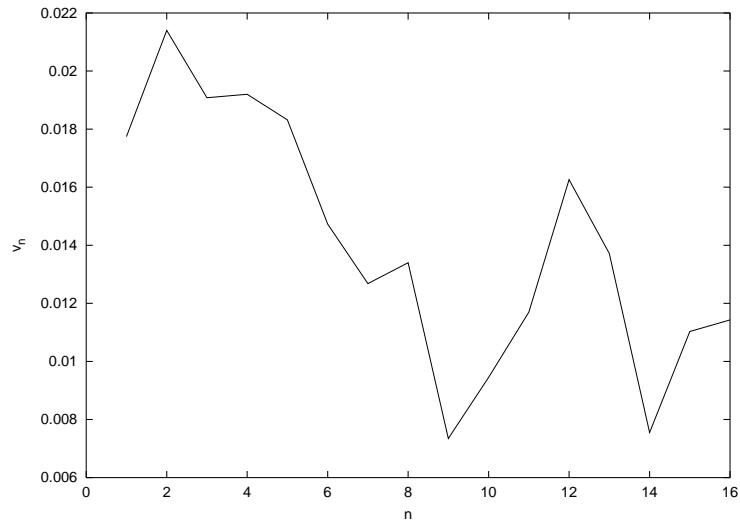,width=10 cm}
\caption{Flow coefficients $v_n$ for pp collisions at 7 TeV
with average multiplicity}
\label{fig7}
\end{figure}

\begin{figure}
\hspace*{2.5 cm}
\epsfig{file=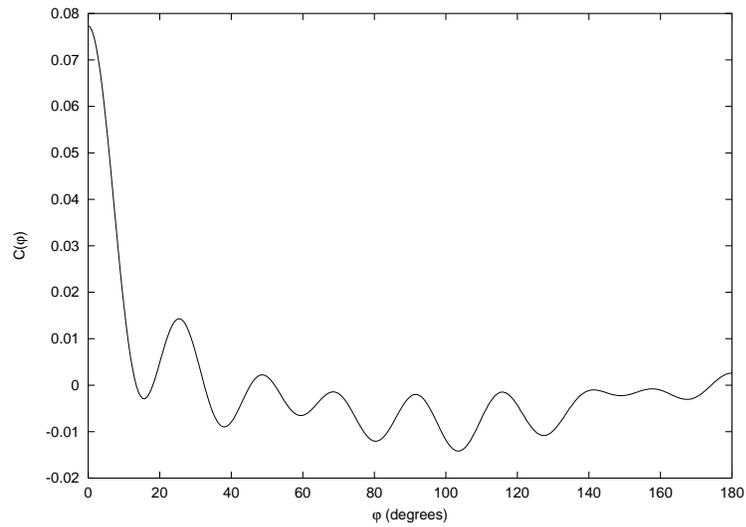,width=10 cm}
\caption{Correlation coeficient $C(\phi)$ for pp
collisions at 7 TeV with average multiplicity}
\label{fig8}
\end{figure}

In the colour string approach proton-proton collisions are described
quite similarly to other hadronic processes. So we appplied
our Monte-Carlo simulations to calculate the flow coefficients in
proton-proton collisions. From  ~\cite{BDMP}
one can conclude that at 62.4, 200 GeV and 7 TeV the average number of
formed strings is 3, 4 and 9 respectively. To describe the impact
parameter dependence
we have assumed the distribution of hadronic matter in the proton to
be a Gaussian with radius 0.8 fm. Calculations show that for such small
number of strings fluctuations are quite strong, so that reliable
results can be obtained after no less that 1000 simulations,
in contrast to the AA case where the resuts are stabilized already at 100
simulations. Our results for the mentioned three energies
are quite similar. So we present them only for the LHC energy of 7
TeV. The maximal number of strings corresponding to the
average one is found to be 18. In Fig. \ref{fig7} we show the
coefficients $v_n$
and in Fig. \ref{fig8} the correlation coefficient $C(\phi)$
averaged over centralitides.
All the $\phi$ dependence is collimated to quite small angles
$\phi\leq 10^0$. Note that
in this case the constant term $DA/<A>^2$ dropped in Fig. \ref{fig8}
is of the order unity, so that the ridge turns out to be only a small
ripple against a constant background.

\begin{figure}
\hspace*{2.5 cm}
\epsfig{file=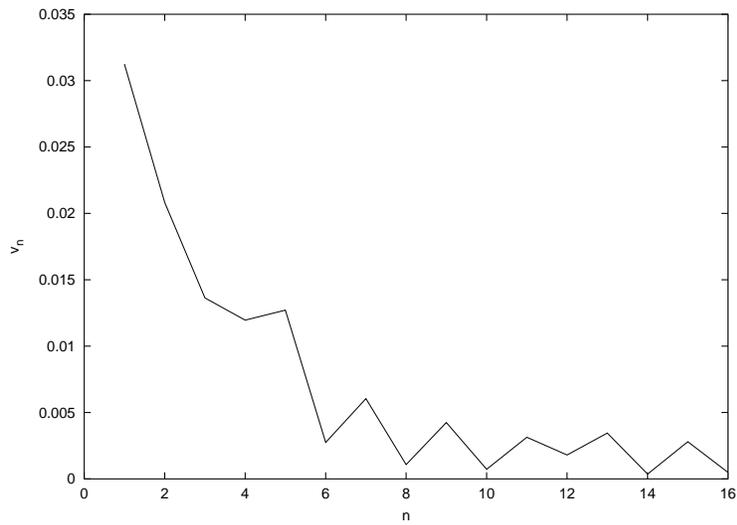,width=10 cm}
\caption{Flow coefficients $v_n$ for pp collisions at 7 TeV
with triple multiplicity}
\label{fig9}
\end{figure}

\begin{figure}
\hspace*{2.5 cm}
\epsfig{file=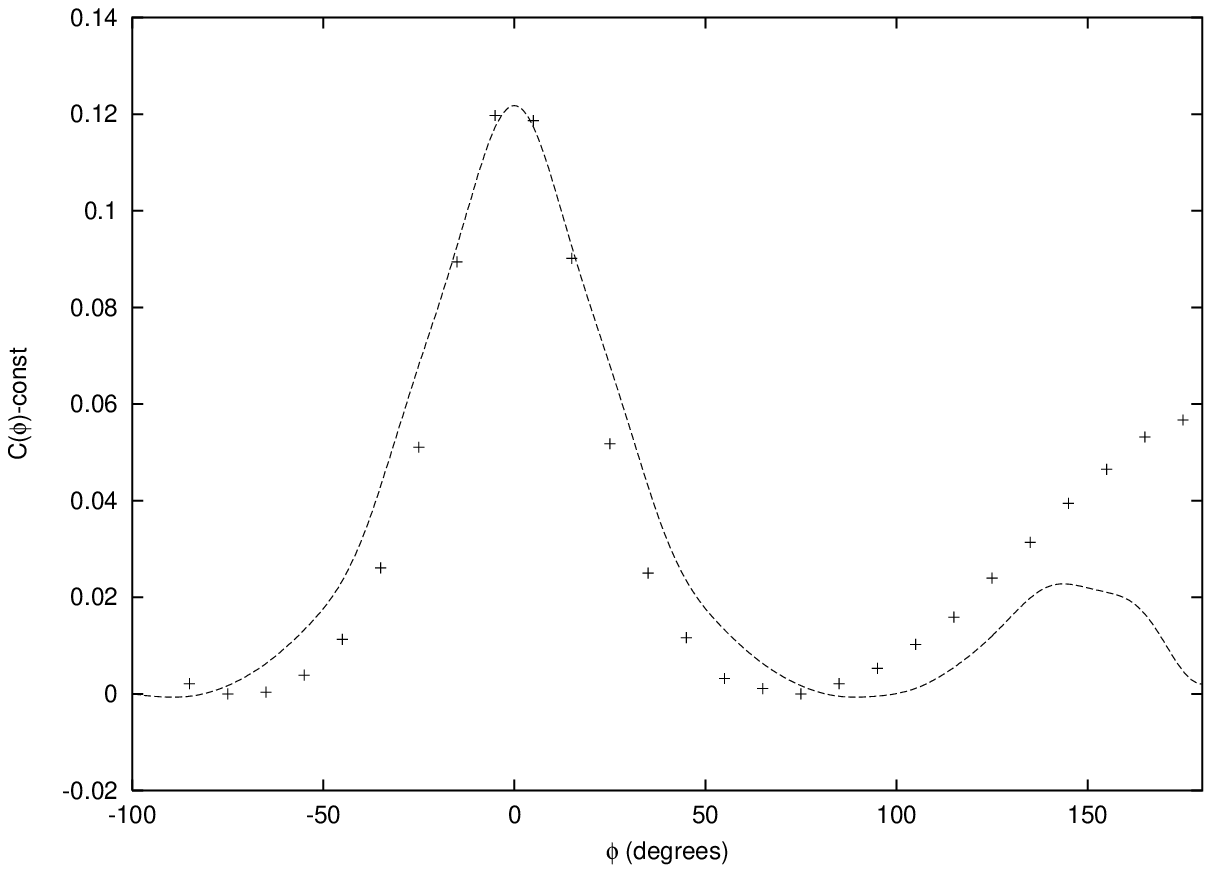,width=10 cm}
\caption{Correlation coeficient $C(\phi)$ for pp
collisions at 7 TeV with triple multiplicity
compared to the the experimental data from ~\cite{CMS}}
(with the ZYAM procedure at positive $\phi$)
\label{fig10}
\end{figure}

Following the experimental observations we studied a rare case in which
the multiplicity is three or more times greater than the average.
The maximal number of string is then found to be 50. The resulting
flow coefficients $v_n$ and correlation coefficients $C(\phi)$ are
shown in Figs. \ref{fig9} and \ref{fig10} respectively. In the latter
figure we compare our results with the experimental data from
 ~\cite{CMS}. As one observes
the correlation coefficient becomes quite similar to the one
in AA collisions in good agreement with the experimenal findings. Still
the dropped constant term is again of the order unity, so that the
ridge stands on a large constant pedestal.

\section{Conclusions}
We have analyzed the possibilty to obtain the long range
azimuthal-rapidity
correlations (ridge) in the framework of the colour string model
with fusion and percolation. An important ingredient in
this approach is  anisotropy of the string
emission spectra in the azimuthal direction which follows
from quenching of the emitted partons in the strong colour field
inside the string ~\cite{ref21}. It is found that one cannot find
such correlations in a single event, since they are severely damped with
the growth of the rapidity difference. Ridge can be obtained only from a
superposition of many events with different numbers and types of strings.
The form of ridge as a function of azimuthal difference  is characterized
by coefficients $w_n$ which are generally different
from the flow coefficients $v_n$ squared and coincide with them only
under certain approximations. We have performed detailed Monte-Carlo
simulations to
find coefficients $w_n$ with $n\leq 16$ and found the ridge correlations
in AA, pA, pp collisions at RHIC and LHC energies. The only adjustable
parameter was taken from comparision with the experimental data for the
elliptic flow coefficient $v_2$ for Au-Au minimum bias collisions at RHIC.
We have confirmed that ridge appears in pp collisions only for events
with an abnormally high multiplicity.

Comparing  with the experimental data at RHIC and LHC
we found a good qualitative agreement for angular correlations in
all cases. As to the quantative agreement it has been found
to be quite good for pp and pPb collisions at 7 and 5.02 TeV,
respectively.
The agreement for AA collisions turned out to be
reasonably good on the near side and on the away side but
worse at intermediate angles where our predictions lie considerably
above the data and the minimum is shfted to  larger angles.
The technical explanantion for this may be related to smaller values for
higher flow coefficients as can be seen in Fig \ref{fig6b}.
As mentioned the reason for this probably can be traced to simplifications
in our Monte-Carlo simulations done to make calculations feasible.
A part of fluctuations were eliminated, which may have lead to non-neglegible
effects especially pronounced in AA collisions.
In any case it is remarkable that the ridge structure in pp, pA and AA collisions can be understood in a unified picture, at least
qualitatively.

Possible refinement of our picture consists of additionally
taking into account harder events, which include jet production and
high-mass diffraction events.
This problem is postponed for future studies.

\section{Acknowledgements}
M.A.B. and V.V.V. have benefited from
grants RFFI 12-02-00356-a and SPbSU 11.38.197.2014 of RUSSIA,
which partially supported this work. C.P. was supported by the project
2011-22776, Consolider of the Ministry of Economy of SPAIN,
Xunta of Galicia and FEDER funds.


\end{document}